\documentclass[lettersize,journal]{IEEEtran}
\usepackage{amsmath, amsthm, amsfonts, amssymb}
\usepackage{algorithmic}
\usepackage{algorithm}
\usepackage{array}
\usepackage{xcolor}
\usepackage[caption=false,font=normalsize,labelfont=sf,textfont=sf]{subfig}
\usepackage{textcomp}
\usepackage{stfloats}
\usepackage{url}
\usepackage{verbatim}
\usepackage{graphicx}
\usepackage{cite}
\usepackage{bm}
\usepackage{booktabs}
\usepackage{caption}

\begin{document}

\title{Age-of-Information Dependent Random Access for Periodic Updating}

\author{Yuqing Zhu, Yiwen Zhu, Aoyu Gong, Yan Lin, and Yijin Zhang
\thanks{This work was supported in part by the National Natural Science Foundation of China under Grant 62071236, and in part by the Open Research Fund of State Key Laboratory of Integrated Services Networks, Xidian University, under Grant ISN22-14. 
}
\thanks{Yuqing Zhu and Yiijn Zhang are with the School of Electronic and Optical Engineering, Nanjing University of Science and Technology, Nanjing 210094, China, and also with the State Key Laboratory of Integrated Services Networks, Xidian University, Xian 710071, China (e-mail: yuqing.zhu@njust.edu.cn; yijin.zhang@gmail.com).}
\thanks{Yiwen Zhu, Yan Lin are with the School of Electronic and Optical Engineering, Nanjing University of Science and Technology, Nanjing 210094, China (e-mail: zyw@njust.edu.cn; yanlin@njust.edu.cn).}
\thanks{A. Gong is with the School of Computer and Communication Sciences, \'Ecole Polytechnique F\'ed\'erale de Lausanne, Lausanne 1015, Switzerland (e-mail: aoyu.gong@epfl.ch).}
}



\maketitle

\begin{abstract}
This paper considers an uplink Internet of Things system with synchronous periodic traffic, where multiple devices generate their status updates at the beginning of each global frame and attempt to send them to a common access point.
To achieve a low network-wide age of information (AoI) in an easily implementable manner, we require each device to adopt an age-dependent random access protocol, i.e., to transmit with a certain probability only when its corresponding AoI reaches a certain threshold.
We analyze the time-average expected AoI by a multi-layer Markov model where an external infinite-horizon Markov chain manages the jumps between the beginnings of frames, while two internal finite-horizon Markov chains manage the evolution during an arbitrary frame for different cases.
Simulation results verify the accuracy of the modeling and the AoI advantage over age-independent schemes.
\end{abstract}


\section{Introduction}
Uplink Internet of Things (IoT) systems have witnessed a rapid growth in real-time
services~\cite{8472907}, such as emergency surveillance, target tracking, and process control.
However, the timeliness requirements of status updating in these services \emph{cannot} be characterized adequately by conventional performance metrics (e.g. throughput and delay).
As such, age of information (AoI) has been introduced in~\cite{2011Minimizing} to measure the difference between the current time and the generation time of the currently newest received update at the destination. 
A fundamental issue in designing access protocols for optimizing AoI is that under a certain traffic pattern, how to schedule short status packets of a large population of devices in an easily implementable manner to achieve a low network-wide AoI.

Prior works~\cite{Gong2020Globecom,Sun2020TCOM,Maatouk2021TWC,Li2022TON} designed optimal or near-optimal policies under various traffic patterns in a centralized manner, which may be impractical for low-cost IoT devices due to a huge amount of signaling overhead for coordination.
To overcome this limited applicability, \cite{Yates2017ISIT,Talak2018SPAWC,Munari2021CL} designed stationary randomized policies of slotted ALOHA, where each device adopts some fixed transmission probability, under the generate-at-will traffic.
\cite{Kosta2019JCN,bae2022age} further considered designing such policies under Bernoulli traffic and synchronous periodic traffic, respectively. 
\cite{bae2022age} also proposed to allow each device to adjust the transmission probability according to the contention level.
However, these policies do not allow each device to utilize the local age knowledge to dynamically adjust the transmission probability, which may lead to AoI performance loss.
Obviously, a successful update from a device with a larger corresponding age would contribute more to reducing the network-wide AoI.
With this observation,~\cite{Atabay2020INFOCOM,chen2020age,Yavascan2021JSAC} investigated age-dependent random access (ADRA), where each device adopts a fixed transmit probability only when its corresponding age reaches a fixed threshold, under the generate-at-will traffic.
This work was extended in~\cite{Chen2022TIT} to consider Bernoulli traffic and a more general ADRA scheme where each device adopts a dynamic transmission probability if its corresponding age gain reaches a fixed or dynamic threshold.
To the best of the author's knowledge, there is no previously known study on ADRA under periodic traffic.

To fill the gap in this field, this paper aims to provide an analytical modeling approach of ADRA under synchronous periodic traffic.
The technical difficulty behind is to consider the mutual impact of periodic update generation and ADRA behavior in modeling, which is overcome by a multi-layer Markov model.
Here an external infinite-horizon Markov chain manages the jumps between the beginnings of frames, while two internal finite-horizon Markov chains manage the evolution during an arbitrary frame for different cases.
Simulation results are presented to demonstrate the accuracy of the analytical modeling and the AoI advantage over the age-independent random access (AIRA) schemes studied in~\cite{bae2022age}.
Note that synchronous periodic traffic is common in closed-loop process control where multiple sensor nodes associated to a process are required to measure the plant outputs and validate the event conditions synchronously and periodically~\cite{Fu2018TAC}.

\section{System Model}  \label{sec:SystemModel}
Consider a globally-synchronized uplink IoT system consisting of a common access point (AP) and $N$ devices, indexed by $\{1,2,...,N\}$.
As shown in Fig.~1, the global channel time is divided into frames (indexed from frame 0), each of which consists of $D$ consecutive slots.
The slots in frame $m$ are indexed from slot $mD$ to $(m+1)D-1$.
At the beginning of each frame, each device synchronously generates a single-slot update and does not generate updates at other time points.
An update is assumed to be successfully transmitted only if its transmission does not overlap with other transmissions on the channel.
After a successful reception of an update of a device, the AP immediately sends an acknowledgement (ACK) to the device without errors. 
At the end of each frame, all the updates that have not been successfully transmitted will be discarded.

\begin{figure}[!ht]
	\centering
	\includegraphics[width=3in]{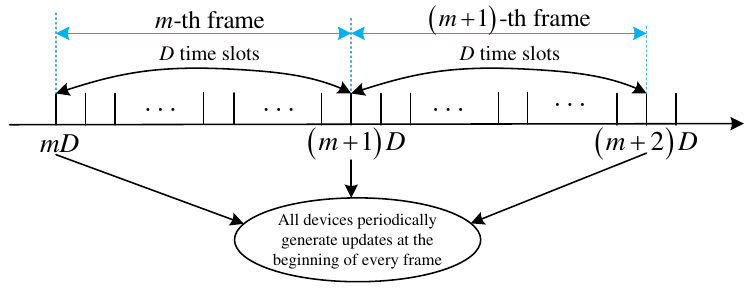}
	\caption{Time Structure.}
	\label{fig:TimeStructure}
\end{figure}

For each device $n$, its instantaneous AoI at the beginning of slot $t$, denoted by $\Delta_{n,t}$, is defined as the number of slots elapsed since the generation moment of its most recently successfully transmitted update.
When slot $t$ is in frame $m$, the evolution of $\Delta_{n,t}$ with $\Delta_{n,0}=0$ can be expressed as
\begin{equation}
	\label{eq:Evolution}
	{\Delta_{n,t+1}} =
	\begin{cases}
		{t+1-mD,}&{\text{if the $m$-th update has been}}\\ &\text{successfully transmitted,} \\
		{\Delta_{n,t}+1,}&{\text{otherwise.}}
	\end{cases}
\end{equation}
The average AoI of device $n$, denoted by ${\Delta_n}$, is given by
\begin{equation}
\label{deqn_ex2a}
    {\Delta_n}=\lim_{T\to\infty}\frac{1}{T}\sum_{t=0}^{T}\Delta_{n,t}. 
\end{equation}

Owing to the ACK mechanism, each device $n$ is able to be aware of the value of $\Delta_{n,t}$ at the beginning of slot $t$.
Then, following~\cite{chen2020age}, we require each device $n$ to deliver its updates according to the following ADRA protocol, that is,
\begin{enumerate}
\item keeps silent at slot $t$ if $\Delta_{n,t}< \delta$ where the age threshold $\delta$ can be an arbitrary non-negative integer.
\item otherwise transmits at the beginning of slot $t$ with the probability $0<p_t\leq1$.
\end{enumerate}
Following~\cite{bae2022age}, we further consider two different settings for determining $p_t$:
\begin{enumerate}
\item the non-adaptive setting: $p_t$ takes a fixed value in $(0,1]$,
\item and the adaptive setting: $p_t=1/u_t$ where $u_t$ denotes the number of contending devices at slot $t$. Note that the value of $u_t$ can be provided by the AP through the control channel, however, how the AP obtains such information is beyond the scope of this work.
\end{enumerate}
\section{Analytical Modeling}  \label{sec:Analytical Modeling}
The symmetric scenario described in Section~\ref{sec:SystemModel} allows us to analyze the average AoI of an arbitrary tagged device to represent the network-wide average AoI. 
We adopt a multi-layer Markov model, where the external layer manages the jumps between the beginnings of frames, while the internal layer manages the evolution during an arbitrary frame.
For analysis simplicity, we omit the device index, identify a slot $t$ by the tuple $(m, h)$ where $m=\lfloor t/D \rfloor$ and $h=t-mD$, and write the age threshold $\delta =\lambda D+\varepsilon$ where $\lambda\in \mathbb{N}$ and $\varepsilon\in \{0,1,\ldots,D-1\}$.
So $u_t$ and $p_t$ can be rewritten as $u_{m,h}$ and $p_{m,h}$, respectively.

\subsection{External Layer} \label{sec:GeneralIdea}
We first present a mathematical tool to model the external layer.
Consider a state process $\bm{X} \triangleq \{X_m, m \in \mathbb{N}\}$ with the initial state $X_0 = 0$, where $X_m$ denotes the instantaneous AoI of the tagged device at the beginning of frame $m$.
By Eq.~\eqref{eq:Evolution}, the evolution of $X_m$ can be expressed as
\begin{equation}
    \label{eq:deqn_Wj}
	X_{m+1} =
	\begin{cases}
		D,    & \text{if the $m$-th update has been} \\
                  & \text{successfully transmitted,} \\
		X_m + D,  & \text{otherwise.}
	\end{cases}
\end{equation}
By Eq.~\eqref{eq:deqn_Wj}, we observe that the transition to the next state in $\bm{X}$ depends only on the present state and not on the previous states.
So, $\bm{X}$ can be viewed as an external discrete-time Markov chain (DTMC) with the infinite state space $\mathcal{X}\triangleq\{0,D,2D,\ldots\}$.

For an arbitrary frame $m$ with $X_m = lD$, let $\alpha_{l,h}$ and $\beta_{l}$ denote the probabilities that the tagged device transmits its $m$-th update successfully in slot $(m,h)$ and in frame $m$, respectively.
Obviously, $\beta_{l} = \sum_{h = 0}^{D-1} \alpha_{l,h}$.
So, the state transition probabilities of $\bm{X}$ can be obtained as
\begin{align}
	R_{x,x'}& \triangleq \text{Pr}(X_{m+1} = x' \mid X_{m} = x) \notag \\
        &=
	\begin{cases} 
		\beta_{l},      & \text{if } x' = D,\, x = lD, \\
		1 - \beta_{l},  & \text{if } x' = (l + 1)D,\, x = lD, \\
            0,              & \text{otherwise.}
	\end{cases}
\end{align}
We discuss possible values of $\alpha_{l,h}$ and $\beta_{l}$ based on the values of $l$, $\lambda$, and $\varepsilon$.
(i) When $l < \lambda$, the tagged device always keeps silent in frame $m$. 
So, we obtain $\alpha_{l,h} = 0$ for $l<\lambda$ and $0\leq h \leq D-1$. 
(ii) When $l = \lambda$, the tagged device keeps silent at slot $(m,h)$ for each $0 \leq h < \varepsilon$, and transmits its $m$-th update with probability $p_{m,h}$ at the beginning of slot $(m,h)$ when $\varepsilon \leq h \leq  D-1$ until the successful transmission.
So, we obtain $\alpha_{\lambda,h} = 0$ for $0\leq h < \varepsilon$ and $0 < \alpha_{\lambda,h} \leq 1$ for $\varepsilon \leq h \leq D-1$.
(iii) When $l > \lambda$, the tagged device transmits its $m$-th update with probability $p_{m,h}$ at the beginning of slot $(m,h)$ when $0\leq h \leq D-1$ until the successful transmission. 
Note that the tagged device behaves the same during frame $m$ regardless of the value of $X_m$ when $X_m > \lambda D$.
So, we write $\alpha_{l,h} = \alpha_{\lambda^+,h}$ and $\beta_{l} = \sum_{h = 0}^{D-1} \alpha_{\lambda^+,h} = \beta_{\lambda^+}$ for all $l>\lambda$. 

\begin{figure}[!ht]
	\centering
	\includegraphics[width=\linewidth]{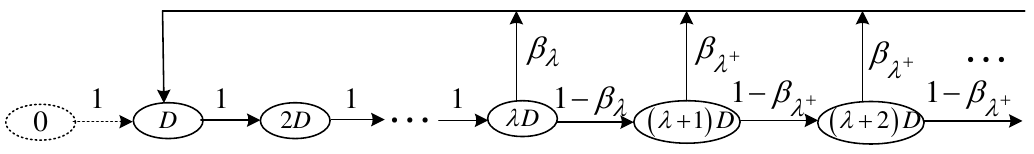}
	\caption{The external DTMC $\bm{X}$.}
	\label{fig:DTMC}
\end{figure}
 
As shown in Fig.~\ref{fig:DTMC}, the state $0$ in $\bm{X}$ is a transient state only occurring when $m = 0$, while the remaining states are all in the same recurrent class.  
As $m$ increases, $\bm{X}$ will get absorbed in the recurrent class and it will stay there forever. 
Let $\pi_{lD}$ denote the steady-state probability of $\bm{X}$ staying at state $lD$ for $l\in\mathbb{Z^+}$.
By the state transition probabilities as shown in~Fig.~\ref{fig:DTMC} and the balance equation $\sum_{l=1}^{\infty}\pi_{lD}=1$, we can obtain	
\begin{align}
     \label{eq:steady-state probabilities}
		{\pi_{lD}}=\begin{cases}
			\cfrac{1}{\lambda + (1-\beta_{\lambda}) / \beta_{\lambda^+}},
			&{\text{if }l<\lambda+1},\\
			\cfrac{1-\beta_{\lambda}}{\lambda + (1-\beta_{\lambda}) / \beta_{\lambda^+}},&{\text{if }l=\lambda+1},\\
			\cfrac{(1-\beta_{\lambda})(1-\beta_{\lambda^+})^{l-\lambda-1}}{\lambda + (1-\beta_{\lambda}) / \beta_{\lambda^+}},&{\text{otherwise}}.
		\end{cases}
\end{align}

Then, for different states in $\bm{X}$, we consider the following two cases for evaluating the average AoI of the tagged device during an arbitrary frame $m$ with $X_m=lD$.

\underline{\emph{Case 1}:} The tagged device transmits its $m$-th update successfully at slot $(m,h)$ given $X_m=lD$.
Let ${\Delta_{l,h}}$ denote the average AoI of the tagged device during frame $m$ when this event occurs. 
We have
\begin{align}
	{\Delta_{l,h}}&=\frac{1}{D}\Big(\sum_{j=0}^{h}\big(lD+j\big)+\sum_{j=h+1}^{D-1}j\Big)  \notag \\
         &=l(h+1)+\frac{D-1}{2}.  \label{eq:Delta1h}
\end{align}

\underline{\emph{Case 2}:} The tagged device fails to transmit its $m$-th update successfully during frame $m$ given $X_m=lD$. 
Let ${\Delta_{l,*}}$ denote the average AoI of the tagged device during frame $m$ when this event occurs. 
We have
\begin{align}
	{\Delta_{l,*}}=\frac{\sum_{h=0}^{D-1}\big(lD+h\big)}{D}=lD+\frac{D-1}{2}.  \label{eq:Delta1*}
\end{align}

Based on the DTMC $\bm{X}$ and Eqs.~\eqref{eq:Delta1h},~\eqref{eq:Delta1*}, the average AoI for an arbitrary device $n$ can be derived as
\begin{align}
  {\Delta_n}
   &=\sum_{l=1}^{\infty}\pi_{lD}\Big(\sum_{h=0}^{D-1}\alpha_{l,h}{\Delta_{l,h}}+(1-\beta_{l}) \Delta_{l,*} \Big) \notag \\
   &=\sum_{l=1}^{\lambda-1}\pi_{lD}\Delta_{l,*} +\pi_{\lambda D}\Big(\sum_{h=\varepsilon}^{D-1}\alpha_{\lambda,h} {\Delta_{\lambda,h}} +\big(1-\beta_{\lambda}\big) \Delta_{\lambda,*}\Big)  \notag\\
	          &\quad+\sum_{l=\lambda+1}^{\infty}\pi_{lD}\Big(\sum_{h=0}^{D-1}\alpha_{\lambda^+,h}{\Delta_{l,h}}
	          + (1-\beta_{\lambda^+})\Delta_{l,*}\Big).     \label{eq:Delta derive}
\end{align}

In the following, we will derive $\alpha_{\lambda,h}$ and $\alpha_{\lambda^+,h}$, which are necessary to derive ${\Delta_n}$ based on Eqs.~\eqref{eq:steady-state probabilities}--\eqref{eq:Delta derive}.

\subsection{Internal Layer to Evaluate $\alpha_{\lambda,h}$}
We now propose the internal layer to evaluate $\alpha_{\lambda,h}$.

Note that whether the tagged device can transmit its $m$-th update successfully depends on the behaviors of all contending devices during frame $m$. 
Let two random variables $S_1$, $S_2$ denote the numbers of other devices (except the tagged device) whose age equals $\lambda D$ and exceeds $\lambda D$ at the beginning of an arbitrary frame $m$, respectively, and let $\chi_{s_1,s_2}$ denote the joint probability mass function of $S_1=s_1$ and $S_2=s_2$.

Let $\alpha_{\lambda,h,s_1,s_2}$ denote the probability that the tagged device transmits its $m$-th update successfully at slot $(m,h)$ when $X_m=\lambda D$, $S_1=s_1$, and $S_2=s_2$. 
Then we have
\begin{align}
	\alpha_{\lambda,h}
			& =\sum_{s_1=0}^{N-1}\sum_{s_2=0}^{N-1-s_1}
				\chi_{s_1,s_2}\alpha_{\lambda,h,s_1,s_2}, \label{eq:alpha_kh}
\end{align}
where
\begin{align}
&\chi_{s_1,s_2}=\binom{N-1}{s_1}\binom{N-1-s_1}{s_2} \notag \\
 & \quad\quad\quad\quad  \times \pi_{\lambda D}^{s_1}\Big(\sum_{l=\lambda+1}^{\infty}\pi_{lD}\Big)^{s_2} \Big(\sum_{l=1}^{\lambda-1}\pi_{lD}\Big)^{N-1-s_1-s_2} \notag \\ 
 &=\frac{(N-1)!}{s_1! s_2! (N - 1 - s_1 - s_2)!} \notag \\
        & \quad\quad\quad \times \pi_{\lambda D}^{s_1} \bigg(\sum_{l=\lambda+1}^{\infty}\pi_{lD}\bigg)^{s_2} \bigg(\sum_{l=1}^{\lambda-1}\pi_{lD}\bigg)^{N-1-s_1-s_2}. \label{eq:Pr_s1s2}
\end{align}

\begin{figure}[!ht]
	\centering
	\includegraphics[width=3.5in]{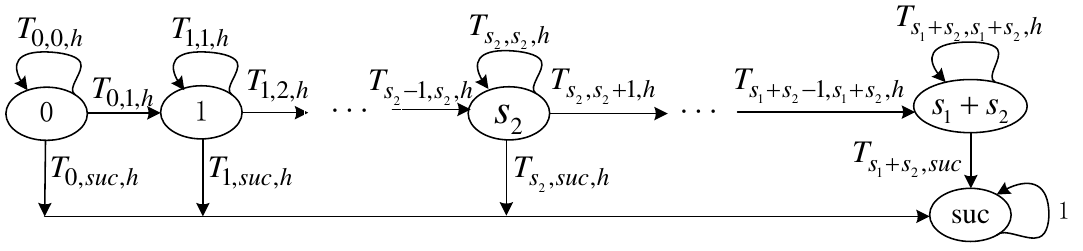}
	\caption{The internal absorbing DTMC $\bm{Y}$.}
	\label{fig:DTMC Y}
\end{figure}

Define $\bm{Y} \triangleq \{Y_h, h=0,1,\ldots,D\}$ as a non-homogeneous absorbing DTMC with the finite state space $\mathcal{Y}\triangleq\{0,1,\ldots,s_1+s_2,suc\}$, as shown in Fig.~\ref{fig:DTMC Y}.
The states $Y_h=y$ with $0\leq y \leq s_1+s_2$ are transient states indicating that the tagged device has not transmitted its $m$-th update successfully while other $y$ devices have transmitted their $m$-th updates successfully at the beginning of slot $(m,h)$ when $X_m=\lambda D$.
The state $suc$ is an absorbing state indicating that the tagged device has transmitted its $m$-th update successfully at the beginning of slot $(m,h)$ when $X_m=\lambda D$. 
For convenience, $h=D$ is used to denote the beginning of slot $(m+1,0)$ here. 
Denote the time-varying state transition probabilities in $\bm{Y}$ by 
\begin{equation}
T_{y,y',h} \triangleq \mathrm{Pr} (Y_{h+1}=y' \mid Y_{h}=y)   \label{eq:tran0}
\end{equation}
for $0 \leq h \leq D-1$, $y,y' \in \mathcal{Y}$.
We consider the following two cases for determining $T_{y,y',h}$ based on the values of $h$ and $\varepsilon$.

\underline{{\emph{Case 1:}}}
When $0\leq h < \varepsilon$, the tagged device keeps silent at slot $(m,h)$.
So, we have
\begin{align} \label{eq:tran1}
    T_{y,y',h}= 
    \begin{cases}
        & 1 - (s_2-y)p_{m,h}\big(1-p_{m,h}\big)^{s_2-y-1},\\
        & \quad\quad \text{if } 0\le y\le s_2-1, y'=y, \\
        & (s_2-y)p_{m,h}\big(1-p_{m,h}\big)^{s_2-y-1}, \\
        & \quad\quad \text{if } 0\le y\le s_2 - 1, y'=y+1, \\
        & 1, \quad \text{if } y'=y = s_2 \text{ or } y'=y = suc, \\
        & 0, \quad \text{otherwise},
    \end{cases}
\end{align}
for all $y,y'\in \mathcal{Y}$.
Here we have $u_{m,h}=s_2-y$.

\underline{{\emph{Case 2:}}}
When $\varepsilon \leq h \leq D-1$, the tagged device transmits its $m$-th update with probability $p_{m,h}$ at the beginning of slot $(m,h)$ until the successful transmission.
So, we have
\begin{align}\label{eq:tran2}
    T_{y,y',h} = 
    \begin{cases} 
        & 1 - (s_1+s_2+1-y)p_{m,h}\big(1-p_{m,h}\big)^{s_1+s_2-y},\\
        & \quad\quad\quad\quad\quad \text{if } 0\le y\le s_1+s_2, y'=y, \\
        & (s_1+s_2-y)p_{m,h}\big(1-p_{m,h}\big)^{s_1+s_2-y}, \\
        & \quad\quad\quad\quad\quad \text{if } 0\le y\le s_1+s_2 - 1, y'=y+1, \\
        & p_{m,h} \big(1-p_{m,h}\big)^{s_1+s_2-y}, \\
        & \quad\quad\quad\quad\quad \text{if } 0 \leq y \leq s_1+s_2, y'=suc, \\
        & 1, \quad\quad\quad\quad \text{if } y'=y = suc, \\
        & 0,  \quad\quad\quad\quad\text{otherwise},
    \end{cases}
\end{align}
for all $y,y'\in \mathcal{Y}$.
Here we have $u_{m,h}=s_1+s_2+1-y$.

Let $\bm{\varphi}_h$ denote the state vector of $Y_h$, where the $i$-th element corresponds to the state $i-1$ for $1\leq i \leq s_1+s_2+1$ and the last element corresponds to the state $suc$.
Then, given the initial state vector $\bm{\varphi}_0 \triangleq (1,0,0,\ldots,0)$ and the time-varying transition matrix $\bm{T}_h$ based on Eqs.~\eqref{eq:tran0}--\eqref{eq:tran2}, by applying a simple power method, we have
\begin{align}
		\bm{\varphi}_{h+1}&=\bm{\varphi}_0 \prod_{j=0}^h \bm{T}_j,   \\
  \alpha_{\lambda,h,s_1,s_2}&=\bm{\varphi}_{h+1}(s_1+s_2+2)-\bm{\varphi}_{h}(s_1+s_2+2),
  \label{eq:alpha_khs1s2}
 \end{align}
for all $0\leq h\leq D-1$ and $0\leq s_1,s_2 \leq N-1$.

\subsection{Internal Layer to Evaluate $\alpha_{\lambda^+,h}$}
We introduce the internal layer to evaluate $\alpha_{\lambda^+,h}$.

Let $\alpha_{\lambda^+,h,s_1,s_2}$ denote the probability that the tagged device transmits its $m$-th update successfully at slot $(m,h)$ when $X_m > \lambda D$, $S_1=s_1$, and $S_2=s_2$. 
We express $\alpha_{\lambda^+,h}$ as
\begin{align}
	\alpha_{\lambda^+,h}
		& =\sum_{s_1=0}^{N-1}\sum_{s_2=0}^{N-1-s_1}
			\chi_{s_1,s_2}\alpha_{\lambda^+,h,s_1,s_2}, \label{eq:alpha_*h}
\end{align}

Similarly, define $\bm{Z} \triangleq \{Z_h, h=0,1,\ldots,D\}$ as a non-homogeneous absorbing DTMC with the finite state space $\mathcal{Z}\triangleq\{0,1,\ldots,s_1+s_2,suc\}$.
The states $Z_h=z$ with $0\leq z \leq s_1+s_2$ are transient states indicating that the tagged device has not transmitted its $m$-th update successfully while other $z$ devices have transmitted their $m$-th updates successfully at the beginning of slot $(m,h)$ when $X_m>\lambda D$.
The state $suc$ is an absorbing state indicating that the tagged device has transmitted its $m$-th update successfully at the beginning of slot $(m,h)$ when $X_m >\lambda D$. 
Denote the time-varying state transition probabilities in $\bm{Z}$ by 
\begin{equation}
Q_{z,z',h} \triangleq \mathrm{Pr} (Z_{h+1}=z' \mid Z_{h}=z)   \label{eq:tran0Q}
\end{equation}
for $0 \leq h \leq D-1$, $z,z' \in \mathcal{Z}$.
We consider the following two cases for determining $Q_{z,z',h}$ based on the values of $h$ and $\varepsilon$.

\underline{{\emph{Case 1:}}}
When $0\leq h <\varepsilon$, we have
\begin{align} \label{eq:tran1Q}
    Q_{z,z',h} = 
    \begin{cases}
        & 1 - (s_2+1-z)p_{m,h}\big(1-p_{m,h}\big)^{s_2-z},\\
        & \quad\quad\quad\quad \text{if } 0\le z\le s_2,\, z'=z, \\
        & (s_2-z)p_{m,h}\big(1-p_{m,h}\big)^{s_2-z}, \\
        & \quad\quad\quad\quad \text{if } 0\le z\le s_2 - 1,\, z'=z+1, \\
        & p_{m,h} \big(1-p_{m,h}\big)^{s_2-z}, \\
        & \quad\quad\quad\quad \text{if } 0 \leq z \leq s_2,\, z'=suc, \\
        & 1, \quad\quad\quad \text{if } z' = z = suc, \\
        & 0, \quad\quad\quad \text{otherwise}.
    \end{cases}
\end{align}
for all $z,z'\in \mathcal{Z}$.
Here we have $u_{m,h}=s_2+1-z$.

\underline{{\emph{Case 2:}}}
When $\varepsilon \leq h \leq D-1$, we have
\begin{equation} \label{eq:tran2Q}
    Q_{z,z',h}=T_{z,z',h},
\end{equation}
for all $z,z'\in \mathcal{Z}$.

Let $\bm{\theta}_h$ denote the state vector of $Z_h$.
Then, given the initial state vector $\bm{\theta}_0 \triangleq (1,0,0,\ldots,0)$ and the time-varying transition matrix $\bm{Q}_h$ based on Eqs.~\eqref{eq:tran0Q}--\eqref{eq:tran2Q}, we have
\begin{align}
		\bm{\theta}_{h+1}&=\bm{\theta}_0 \prod_{j=0}^h \bm{Q}_j,   \\
  \alpha_{\lambda^+,h,s_1,s_2}&=\bm{\theta}_{h+1}(s_1+s_2+2)-\bm{\theta}_{h}(s_1+s_2+2),
  \label{eq:alpha_*hs1s2}
 \end{align}
for all $0\leq h\leq D-1$ and $0\leq s_1,s_2 \leq N-1$.

\subsection{Evaluation of $\Delta_n$}
Now we are ready to compute $\Delta_n$ by connecting the internal and external layers proposed in previous subsections.

Jointly considering $\beta_{\lambda} = \sum_{h = 0}^{D-1} \alpha_{\lambda,h}$, $\beta_{\lambda^+} = \sum_{h = 0}^{D-1}$ $\alpha_{\lambda^+,h}$, and Eqs.~\eqref{eq:steady-state probabilities},~\eqref{eq:alpha_kh},~\eqref{eq:Pr_s1s2},~\eqref{eq:alpha_*h}, we obtain nonlinear equations for $\beta_{\lambda}$ and $\beta_{\lambda^+}$, which can be solved by numerical methods. 
With the solutions of $\beta_{\lambda}$ and $\beta_{\lambda^+}$, we can then obtain the steady-state probabilities given in Eq.~\eqref{eq:steady-state probabilities} and the values of $\alpha_{\lambda,h}$, $\alpha_{\lambda^+,h}$ given in Eqs.~\eqref{eq:alpha_kh},~\eqref{eq:alpha_*h}. 
Finally, we can obtain the average AoI for an arbitrary device $n$ given in Eq.~\eqref{eq:Delta derive}.

\vspace{2pt}

\noindent \emph{Remark 1:} 
When the period $D=1$ (i.e., the generate-at-will traffic) and $p_t$ takes a fixed value, we note
\begin{align}
	\beta_{\lambda} = \beta_{\lambda^+}=\alpha_{\lambda,0} &= \alpha_{\lambda^+,0} \notag\\
	&=\sum_{s_1=0}^{N-1}\sum_{s_2=0}^{N-1-s_1}\chi_{s_1,s_2}p_{m,0} \big(1-p_{m,0}\big)^{s_1+s_2}\notag\\
	&=p_{m,0}(1-p_{m,0}\sum_{l=\lambda}^{\infty}\pi_{lD})^{N-1}. 
	\end{align}
So our modeling approach is reduced to that in~\cite{chen2020age} for $D=1$. 

\vspace{2pt}

\noindent \emph{Remark 2:} 
When the age threshold $\delta=0$, we note that we can drop the subscript $l$ of $\beta_l$ and have $\pi_{lD}=\beta(1-\beta)^{l-1}$ for $l\in\mathbb{Z^+}$ since $\beta_l$ is independent of $l$.
So, our modeling approach is reduced to that in~\cite{bae2022age} for analyzing AIRA.

\vspace{2pt}

\noindent \emph{Remark 3:} 
For the non-adaptive setting of $p_t$, the age threshold $\delta$ and fixed $p_t$ should be jointly optimized to achieve good performance. 
With some performance loss, one also can choose a reasonable fixed $p_t$ and optimize $\delta$ based on a one-dimensional search.
For the adaptive setting of $p_t$, only $\delta$ needs to be optimized under $p_t=1/u_t$ by a one-dimensional search at the cost of additional overhead for determining $u_t$.


\section{Numerical Results}
We compare the analytical and stimulative results of the proposed ADRA, and then examine the performance advantage over AIRA~\cite{bae2022age} under adaptive and optimal non-adaptive settings of $p_t$ as described in section~\ref{sec:SystemModel}.
The scenarios considered in the simulations are in accordance with the descriptions in Section~II.
Each simulation result is obtained from 10 independent simulation runs with $10^7$ slots in each run.

\begin{figure}[!ht]
	\centering
	\includegraphics[width=3.1in]{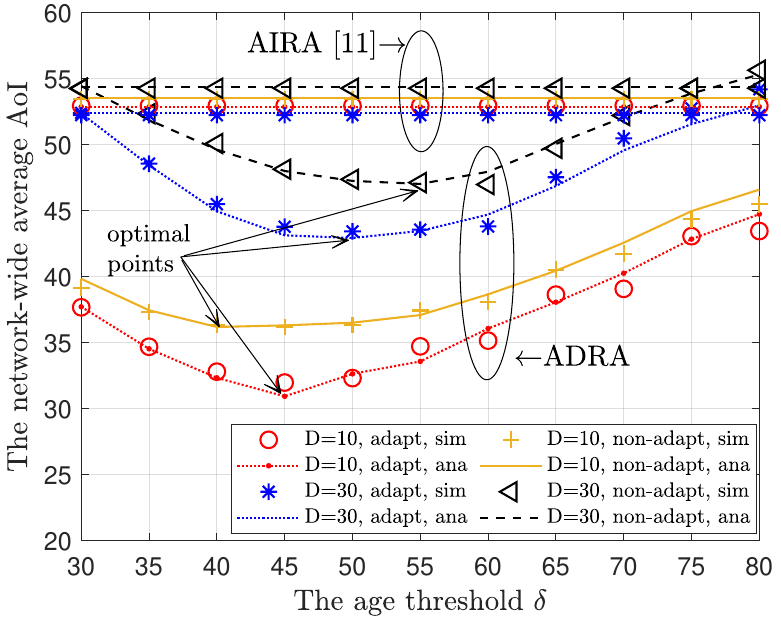}
	\caption{The network-wide average AoI versus the age threshold $\delta$ for $D=10, 30$ and $N = 20$. Here the ADRA is obtained by searching optimal $p_t$ under a given $\delta$.}
	\label{fig:verus delta}
\end{figure}

Fig.~\ref{fig:verus delta} shows the network-wide average AoI as a function of the age threshold $\delta$ when $N=20$ and $D=10, 30$. 
The curves indicate that our analytical model is accurate in all the cases.
We observe that under each setting of $p_t$, the AoI of ARDA always first decreases with $\delta$ and then increases with $\delta$.
This is because small $\delta$ would lead to severe contention while large $\delta$ would lead to few transmission opportunities.
We further observe that optimal ADRA always significantly outperforms the AIRA: $13.44\%$--$34.16\%$ improvement under optimal fixed $p_t$,  $16.85\%$--$39.59\%$ improvement under $p_t=1/u_t$.
These comparisons confirm the effectiveness of introducing $\delta$, which enables uncoordinated devices to contend in a more harmonic way to reduce the network-wide AoI together. 

\begin{figure}[!ht]
	\centering
	\includegraphics[width=3.1in]{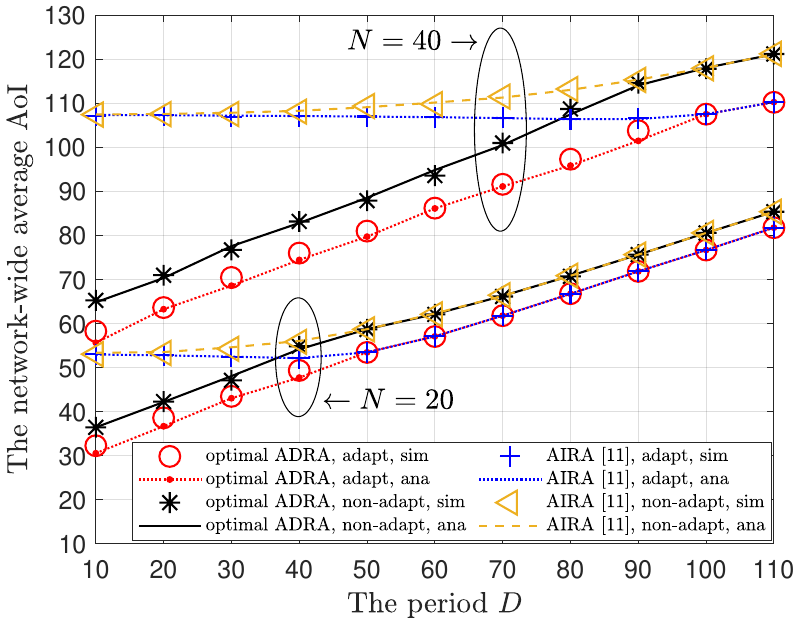}
	\caption{The network-wide average AoI versus the period $D$ for $N=20,40$. 
 }
	\label{fig:verus D}
\end{figure}

Fig.~\ref{fig:verus D} shows the network-wide average AoI as a function of the period $D$ when $N=20, 40$. 
The curves verify the accuracy of our analytical model again.
We observe that the AoI of AIRA first keeps almost the same and then increases with $D$,  
while the AoI of optimal ARDA always increases almost linearly with $D$.
This is because introducing $\delta$ prioritizes those devices with large AoI to transmit with less collisions, which mitigates the severe contention caused by small $D$. 
We also observe that the optimal ADRA always outperforms AIRA: $0\%$--$39.31\%$ improvement under optimal fixed $p_t$,  $0\%$--$45.69\%$ improvement under $p_t=1/u_t$, which becomes smaller as $D$ increases or $N$ decreases.
This is because the effect of introducing $\delta$ to mitigate the contention becomes smaller under larger $D$ or smaller $N$. 

\section{Conclusion}
This paper has developed a multi-layer Markov modeling approach for evaluating the AoI of ADRA in an uplink IoT system with synchronous periodic updating.
An external DTMC is used to manage the jumps between the beginnings of frames, while two internal absorbing DTMCs are used to manage the evolution during an arbitrary frame.
Simulation results validated our theoretical study and confirmed the advantage over AIRA.
The results also revealed that different from AIRA~\cite{bae2022age}, the optimal ADRA utilizes the benefit of short update period on the AoI efficiently, so that shorter update period is preferable to reduce the AoI.


\bibliography{test}
\end{document}